\documentclass[11pt]{scrartcl}

\usepackage[fleqn]{amsmath}
\usepackage{amssymb,amsmath,amsthm}
\usepackage{amsfonts}
\usepackage{xcolor}
\usepackage{lineno,hyperref}
\usepackage[english]{babel}
\usepackage[square,sort,comma,numbers]{natbib}
\usepackage{graphicx}
\usepackage[utf8]{inputenc}

\begin{document}

\title{Approximating the discrete time-cost tradeoff problem with bounded depth}

\author{Siad Daboul \qquad Stephan Held \qquad Jens Vygen}
\date{\small Research Institute for Discrete Mathematics and Hausdorff Center for Mathematics\\
University of Bonn\\
\{daboul, held, vygen\}@dm.uni-bonn.de} %
\maketitle

\begin{abstract}
We revisit the deadline version of the discrete time-cost tradeoff problem for the special case of
bounded depth. Such instances occur for example in VLSI design.
The depth of an instance is the number of jobs in a longest chain and is denoted by $d$.
We prove new upper and lower bounds on the approximability.

First we observe that the problem can be regarded as a special case of finding a minimum-weight vertex cover in
a $d$-partite hypergraph.
Next, we study the natural LP relaxation, which can be solved in polynomial time for fixed $d$
and --- for time-cost tradeoff instances --- up to an arbitrarily small error in general.
Improving on prior work of Lov\'asz and of Aharoni, Holzman and Krivelevich, 
we describe a deterministic algorithm with approximation ratio slightly less than $\frac{d}{2}$
for minimum-weight vertex cover in $d$-partite hypergraphs for fixed $d$ and given $d$-partition.
This is tight and yields also a $\frac{d}{2}$-approximation algorithm for general time-cost tradeoff instances.

We also study the inapproximability 
and show that 
no better approximation ratio than
$\frac{d+2}{4}$ is possible, assuming the Unique Games Conjecture and $\textnormal{P}\neq\textnormal{NP}$. 
This strengthens a result of Svensson \cite{Svensson}, who showed that under the same assumptions no constant-factor
approximation algorithm exists for general time-cost tradeoff instances (of unbounded depth).
Previously, only APX-hardness was known for bounded depth.
\end{abstract}

\newtheorem{thm}{Theorem}
\newtheorem{lem}[thm]{Lemma}
\newtheorem{prop}[thm]{Proposition}
\newtheorem{cor}[thm]{Corollary}
\newtheorem{defn}[thm]{Definition}

\newenvironment{pf}
  {\begin{proof}[Proof]\let\qed\relax}
  {\end{proof}}

\newtheorem*{definition*}{Definition}
\newtheorem{definition}{Definition}

\newcommand*{\QEDB}{\hfill\ensuremath{\blacksquare}}%
\newcommand*{\QED}{\hfill\ensuremath{\square}}%


\section{Introduction}
The (deadline version of the discrete) time-cost tradeoff problem
was introduced in the context of project planning and scheduling more than 60 years ago \cite{Kelley59}.
An instance of the \emph{time-cost tradeoff problem} consists of a finite set $V$ of jobs,
a partial order $(V,\prec)$, a deadline $T>0$,
and for every job $v$ a finite nonempty set $S_v \subseteq \mathbb{R}_{\geq 0}^2$ of time/cost pairs.
An element $(t,c) \in S_v$ corresponds to a possible choice of performing job $v$ with delay $t$ and cost $c$.
The task is to choose a pair $(t_v,c_v) \in S_v$ for each $v \in V$ such that $\sum_{v \in P}t_v \leq T$ for every chain $P$
(equivalently: the jobs can be scheduled within a time interval of length $T$, respecting the precedence constraints),
and the goal is to minimize $\sum_{v \in V} c_v$.

The partial order can be described by an acyclic digraph $G=(V,E)$, where $(v,w) \in E$ if and only if $v \prec w$.
Every chain of jobs corresponds to a path in $G$, and vice versa.

De et al.\ \cite{TctNpHard} proved that this problem is strongly NP-hard.
Indeed, there is an approximation-preserving reduction from vertex cover \cite{NoConstantTct},
which implies that, unless $\textnormal{P}=\textnormal{NP}$, there is no 1.3606-approximation algorithm \cite{VertexCover}.
Assuming the Unique Games Conjecture and $\textnormal{P}\neq\textnormal{NP}$, Svensson \cite{Svensson} could show
that no constant-factor approximation algorithm exists.

Even though the time-cost tradeoff has been extensively studied due to its numerous practical applications,
only few positive results about approximation algorithms are known.
Skutella~\cite{Skutella} described an algorithm that works if all delays are natural numbers in the range $\{0,\ldots,l\}$
and returns an $l$-approximation.
If one is willing to relax the deadline, one can use Skutella's bicriteria approximation algorithm \cite{Skutella}.
For a fixed parameter $0<\mu<1$, it
computes a solution in polynomial time such that the optimum cost is exceeded by a factor of at most $\frac{1}{1-\mu}$
and the deadline $T$ is exceeded by a factor of at most $\frac{1}{\mu}$.
Unfortunately, for many applications, including VLSI design, relaxing the deadline is out of the question.

The instances of the time-cost tradeoff problem that arise in the context of VLSI design
usually have a constant upper bound $d$ on the number of vertices on any path \cite{ByeVlsi}.
This is due to a given target frequency of the chip, which can only be achieved if the logic depth is bounded.
For this important special case, we will describe better approximation algorithms.

The special case $d=2$ reduces to weighted bipartite matching and can thus be solved optimally in polynomial time.
However, already the case $d=3$ is strongly NP-hard \cite{TctNpHard}. The case $d=3$ is even APX-hard, because
De\v{\i}neko and Woeginger~\cite{HardnessTct} devised
an approximation-preserving reduction from vertex cover in cubic graphs
(which is known to be APX-hard \cite{VCC}).

On the other hand, it is easy to obtain a $d$-approximation algorithm: either by applying the
Bar-Yehuda--Even algorithm for set covering \cite{Bye,ByeVlsi} or (for fixed $d$) by simple LP rounding;
see the end of Section \ref{sec:setcoverLP}.

As we will observe in Section \ref{sec:setcoverLP}, the time-cost tradeoff problem with depth $d$
can be viewed as a special case of finding a minimum-weight vertex cover in a $d$-partite hypergraph.
Lov\'asz \cite{LovaszDiss} studied the unweighted case and proved that the natural LP has integrality gap $\frac{d}{2}$.
Aharoni, Holzman and Krivelevich \cite{Aharoni} showed this ratio for more general unweighted hypergraphs by randomly rounding a given LP solution.
Guruswami, Sachdeva and Saket \cite{InapproxHypGraph} proved that approximating the vertex cover problem in $d$-partite hypergraphs
with a better ratio than $\frac{d}{2}-1+\frac{1}{2d}$ is NP-hard, and better than $\frac{d}{2}$ is NP-hard if the Unique Games Conjecture holds.

\section{Results and Outline}

In this paper, we first reduce the time-cost tradeoff problem with depth $d$
to finding a minimum-weight vertex cover in a $d$-partite hypergraph.
Then we simplify and derandomize the LP rounding algorithm of Lov\'asz \cite{LovaszDiss} and Aharoni et al.\ \cite{Aharoni}
and show that it works for general nonnegative weights.
This yields a simple deterministic $\frac{d}{2}$-approximation algorithm for minimum-weight vertex cover
in $d$-partite hypergraphs for fixed $d$, given $d$-partition, and given LP solution.
To obtain a $\frac{d}{2}$-approximation algorithm for the time-cost tradeoff problem,
we develop a slightly stronger bound for rounding the LP solution, because the vertex cover LP can only be solved approximately (unless $d$ is fixed).
This will imply our first main result:

\begin{thm}
\label{thm:mainub}
There is a polynomial-time $\frac{d}{2}$-approximation algorithm for the time-cost tradeoff problem,
where $d$ denotes the depth of the instance. 
\end{thm}

The algorithm is based on rounding an approximate solution to the vertex cover LP.
The basic idea is quite simple: 
we partition the jobs into levels and carefully choose an individual threshold for every level,
then we accelerate all jobs for which the LP solution is above the threshold of its level.
We get a solution that costs slightly less than $\frac{d}{2}$ times the LP value.
Since the integrality gap is $\frac{d}{2}$ \cite{LovaszDiss,Aharoni}
(even for time-cost tradeoff instances; see Section~\ref{sec:setcoverLP}), this ratio is tight.

The results by \cite{InapproxHypGraph} suggest that this approximation guarantee is essentially best possible
for general instances of the vertex cover problem in $d$-partite hypergraphs.
Still, better algorithms might exist for special cases such as the time-cost tradeoff problem.
However, we show that much better approximation algorithms are unlikely to exist even for time-cost tradeoff instances.
More precisely, we prove:
\begin{thm}
\label{thm:mainlb}
Let  $d \in \mathbb{N}$ with $d \geq 2$ and $\rho<\frac{d+2}{4}$ be constants.
Assuming the Unique Games Conjecture and $\textnormal{P}\neq\textnormal{NP}$,
there is no polynomial-time $\rho$-approximation algorithm for time-cost tradeoff instances with depth $d$.
\end{thm}

This gives strong evidence that our approximation algorithm is best possible up to a factor of 2.
To obtain our inapproximability result, we leverage Svensson's theorem on the hardness of vertex deletion to destroy
long paths in an acyclic digraph \cite{Svensson} and strengthen it to instances of bounded depth
by a novel compression technique.

Section \ref{sec:setcoverLP} introduces the vertex cover LP and explains why the time-cost tradeoff problem with depth $d$
can be viewed as a special case of finding a minimum-weight vertex cover in a $d$-partite hypergraph.
In Section \ref{sec:algorithm} we describe our approximation algorithm, which rounds a solution to this LP.
Then, in Sections \ref{sec:inapprox} and \ref{sec:depthreduction} we prove our inapproximability result.

\section{The vertex cover LP\label{sec:setcoverLP}}

Let us define the \emph{depth} of an instance of the time-cost tradeoff problem to be
the number of jobs in the longest chain in $(V,\prec)$, or equivalently the number of vertices in the longest path 
in the associated acyclic digraph $G=(V,E)$.
We write $n=|V|$, and the depth will be denoted by $d$ throughout this paper.

First, we note that one can restrict attention to instances with a simple structure,
where every job has only two alternatives and the task is to decide which jobs to accelerate.
This has been observed already by Skutella \cite{Skutella}.
The following definition describes the structure that we will work with.

\begin{defn}
An instance $I$ of the time-cost tradeoff problem is called \emph{normalized} if for each job $v \in V$ 
the set of time/cost pairs is of the form $S_v=\{(0,c),(t,0)\}$ for some $c,t \in \mathbb{R}_+ \cup \{\infty\}$.
\end{defn}

In a normalized instance, every job has only two possible ways of being executed.
The slow execution is free and the fast execution has a delay of zero. Therefore, the time-cost tradeoff
problem is equivalent to finding a subset $F \subseteq V$ of jobs that are to be executed fast. 
The objective is to minimize the total cost of jobs in $F$.
Note that for notational convenience we allow one of the alternatives to have infinite delay or cost,
but of course such an alternative can never be chosen in a feasible solution of finite cost, and it could be as well excluded.

We call two instances $I$ and $I'$ of the time-cost tradeoff problem \textit{equivalent} if
any feasible solution to $I$ can be transformed in polynomial time to a feasible solution to $I'$ with the same cost and vice-versa.
We include a proof of Skutella's observation for sake of completeness.

\begin{prop}[Skutella \cite{Skutella}]
\label{prop:precondition}
For any instance $I$ of the time-cost tradeoff problem one can construct an equivalent normalized instance $I'$ of the same depth in polynomial time.
\end{prop}

\begin{pf}
Let $v$ be a job of instance $I$ with $S_v=\{(t_1,c_1),$ $\ldots,(t_r,c_r)\}$.
By sorting and removing dominated pairs, we may assume $t_1 < \ldots < t_r$ and $c_1 > \ldots > c_r$.

To construct $I'$,
we replace $v$ by $r+1$ copies $v_0,v_1,\ldots,v_r$ of $v$, each with the same predecessors and successors.
We define $S_{v_i}:=\{(0,c_i-c_{i+1}),(t_{i+1},0)\}$, where $c_0:= \infty$, $c_{r+1}:=0$, and $t_{r+1}:=\infty$.

As the slow alternatives of the copies $v_i$ have increasing delay in $i$,
an optimum solution always sets consecutive jobs $v_j,v_{j+1},\ldots v_r$ to the fast solution.
As the last slow solution has infinite delay and the first one has infinite cost, $1\leq j \leq r$.
Then the total cost at $v$ is given by $\sum_{i=j}^r \left(c_i-c_{i+1}\right)=c_j-c_{r+1}=c_j$.
As accelerated jobs have delay 0, the longest path through a copy of $v$ is determined by $v_{j-1}$, which has delay $t_j$.

Note that it is easy to convert the corresponding solutions of both instances into each other in polynomial time.
\QED
\end{pf}

The structure of only allowing two execution times per job gives rise to a useful property, as we will now see.
As noted above, for a normalized instance $I$ the solutions correspond to subsets of jobs $F \subseteq V$
to be accelerated.
Consider the clutter $\mathcal{C}$ of inclusion-wise minimal feasible solutions to $I$.
Denote by $\mathcal{B}=\text{bl}(\mathcal{C})$ the blocker of $\mathcal{C}$,
i.e., the clutter over the same ground set $V$ whose members are minimal subsets of jobs that
have nonempty intersection with every element of $\mathcal{C}$.

Let $T > 0$ be the deadline of our normalized time-cost tradeoff instance and $t_v$ denote the slow delay of executing job $v \in V$.
By the properties of a normalized instance, the elements of $\mathcal{B}$ are
the minimal chains $P \subseteq V$ with $\sum_{v \in P}t_v > T$.
The well-known fact that $\text{bl}(\text{bl}(\mathcal{C}))=\mathcal{C}$ \cite{Clutter, ClutterIdent}
immediately implies the next proposition, which also follows from an elementary calculation.

\begin{prop}\label{prop:coverprop}
A set $F \subseteq V$ is a feasible solution to a normalized instance $I$ of the time-cost tradeoff problem if and only if
$P \cap F\neq \emptyset$ for all $P \in \mathcal{B}$.
\qed
\end{prop}

Therefore, our problem is to find a minimum-weight vertex cover in the hypergraph $(V, \mathcal{B})$. If our time-cost tradeoff instance has depth $d$, this hypergraph is $d$-partite\footnote{A hypergraph $(V, \mathcal{B})$ is $d$-partite if there exists a partition $V = V_1\dot{\cup}V_2\dots \dot{\cup}V_d$ such that $|P\cap V_i| \le 1$ for all $P \in \mathcal{B}$ and $i \in \{1,\dots, d\}$. We call $\{V_1,\dots, V_d\}$ a $d$-partition. We do not require the hypergraph to be $d$-uniform.} and a $d$-partition can be computed easily:

\begin{prop}\label{prop:layers}
Given a time-cost tradeoff instance with depth $d$, we can partition the set of jobs in polynomial time
into sets $V_1,\ldots,V_d$ (called \emph{layers})
such that $v \prec w$ implies that $v \in V_i$ and $w \in V_j$ for some $i<j$. Then, $|P\cap V_i| \leq 1$ for all $P \in \mathcal{B}$ and $i=1,\ldots,d$.
\end{prop}

\begin{pf}
Such a partition can be found by constructing the acyclic digraph $G=(V,E)$ with
$(v,w)\in E$ if and only if $v\prec w$ and setting $V_i:=\{v \in V: l(v)=i\}$,
where $l(v)$ denotes the maximum number of vertices in any path in $G$ that ends in $v$.
\QED
\end{pf}

This also leads to a simple description as an integer linear program.
The feasible solutions correspond to the vectors $x \in \{0,1\}^V$ with $\sum_{v \in P}x_v \geq 1$ for all $P \in \mathcal{B}$. 
We consider the following linear programming relaxation, which we call the \emph{vertex cover LP}:

\begin{alignat}{2}
  & \text{minimize: } & & \phantom{..}\sum_{v \in V} c_v \cdot x_v  \nonumber \\
  & \text{subject to: } & & \phantom{..}\begin{aligned}[t]
                     & \sum_{v \in P} x_v\geq 1& ~~\text{for all } P \in \mathcal{B} \label{knapsacklp} \\[0ex]
                    & x_v\geq 0  & ~~\text{for all } v \in V.
                \end{aligned}
\end{alignat}

Let $\text{LP}$ denote the value of this linear program (for a given instance).
Unless P=NP, one cannot solve this linear program exactly in polynomial time:

\begin{prop}
If the vertex cover LP \eqref{knapsacklp} can be solved in polynomial time
for normalized time-cost tradeoff instances, then $\textnormal{P}=\textnormal{NP}$.
\end{prop}

\begin{pf}
By the equivalence of optimization and separation \cite{GrotschelLovaszSchrijver},
it suffices to show that the separation problem is NP-hard.
In fact, we show that deciding whether a given vector $x$ is infeasible for a given instance is NP-complete.
To this end, we transform the \textsc{Partition} problem, which is well known to be NP-complete:
given a list $a_1,\ldots,a_n$ of positive integers, is there a subset $I\subseteq\{1,\ldots,n\}$ with
$\sum_{i\in I}a_i=\sum_{i\notin I}a_i$?
Given an instance $a_1,\ldots,a_n$ of \textsc{Partition}, construct a time-cost tradeoff instance
with $2n$ jobs $v_{ij}$ ($i=1,\ldots,n$, $j=0,1$), where $v_{ij}\prec v_{i'j'}$ whenever $i<i'$.
The fast execution time is 0 for all jobs, and the slow execution time is also 0 for $v_{i0}$ but $a_i$ for $v_{i1}$.
The deadline is $T:=\frac{A-1}{2}$, where $A=\sum_{i=1}^n a_i$.
Let $x_{v_{i0}}:=0$ and $x_{v_{i1}}:=\frac{2a_i}{A+1}$.
Then $x$ is a feasible solution to the LP if for all subsets $I\subseteq\{1,\ldots,n\}$
$\sum_{i\in I}a_i \le T$ or $\sum_{i\in I} x_{v_{i1}}\ge 1$, which means
$\sum_{i\in I}a_i \le \frac{A-1}{2}$ or $\sum_{i\in I}a_i \ge \frac{A+1}{2}$, or equivalently
$\sum_{i\in I}a_i\neq\frac{A}{2}$.
\QED
\end{pf}

However, we can solve the LP up to an arbitrarily small error;
in fact, there is a fully polynomial approximation scheme
(as essentially shown by \cite{KarmarkarKarp}):

\begin{prop}\label{prop:solvelp}
For normalized instances of the time-cost tradeoff problem with bounded depth,
the vertex cover LP \eqref{knapsacklp} can be solved in polynomial time.
For general normalized instances and any given $\epsilon>0$, a feasible solution of cost
at most $(1+\epsilon)\textnormal{LP}$ can be found in time bounded by a
polynomial in $n$ and $\frac{1}{\epsilon}$.
\end{prop}

\begin{pf}
If the depth is bounded by a constant $d$, the number of constraints is bounded by the polynomial $|V|^d$,
so the first statement follows from any polynomial-time linear programming algorithm.

Otherwise, we solve the LP up to a factor $1+\epsilon$ for any given $0<\epsilon\leq 1$ as follows.
Implement an approximate separation oracle by first rounding up
the components of a given vector $x$ to integer multiples of $\frac{\epsilon}{2d}$ and
then applying dynamic programming (similar to the knapsack problem) to check whether
$\sum_{v\in P} \frac{\epsilon}{2d} \lceil\frac{2d x_v}{\epsilon}\rceil \ge 1$ for all $P\in\mathcal{B}$.
This requires $\mathcal{O}(\frac{dn^2}{\epsilon})$ time.

Run the ellipsoid method with this oracle. It computes an optimum solution $x$ to a relaxed linear program, hence with cost at most $\text{LP}$.
Moreover, $(1+\epsilon)x$ is a feasible solution to the original LP \eqref{knapsacklp} because
for every $P\in\mathcal{B}$ we have
$\frac{\epsilon}{2}+\sum_{v\in P} x_v \ge \sum_{v\in P} \bigl( x_v+\frac{\epsilon}{2d} \bigr) \ge 1$,
implying $(1+\epsilon)\sum_{v\in P} x_v \ge (1+\epsilon)(1-\frac{\epsilon}{2}) \ge 1$.
\QED
\end{pf}

\begin{figure}[tb]
\centering
\includegraphics[scale=3.5,clip=false]{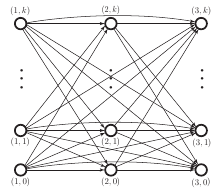}
\caption{
An instance of the $d$-partite hypergraph vertex cover problem given by \cite{Aharoni},
which can also be interpreted as a normalized instance of the time-cost tradeoff problem.
We have $n=d(k+1)$ jobs $V=\{(i,j) \,|\, i=1,\ldots, d,\, j=0,\,\ldots,k\}$ for some $k \in \mathbb{N}$, with
$(i,j)\prec (i',j')$ whenever $i< i'$. The deadline is given by $T=\frac{dk}{2}$.
The slow variant of job $(i,j)$ has delay $j$ without any cost.
By paying a cost of $1$ the delay drops to $0$.
The above figure illustrates this for $d=3$.
One can easily verify that assigning vertex $(i,j)$ a fractional value of $x_{(i,j)}=\frac{j}{T}$ is feasible
with total cost $k+1$. Let $\tau_i$ be the number of vertices in $V_i=\{(i,0), \ldots, (i,k)\}$ that an optimum solution accelerates.
The delay of the slowest job in level $i$ is then at least $k-\tau_i$, and
we conclude that
$\sum_{i=1}^d (k-\tau_i) \leq T$  and hence $\sum_{i=1}^d\tau_i \geq d k -T = \frac{dk}{2}$.
Therefore the integrality gap is at least $\frac{d k/2}{k+1}
 \xrightarrow[k \rightarrow \infty]{}  \frac{d}{2}$.
}
\label{fig:gap}
\end{figure}

We remark that the $d$-partite hypergraph vertex cover instances given by \cite{Aharoni} can be also considered as normalized
instances of the time-cost tradeoff problem; see Figure \ref{fig:gap}. This  shows that the integrality gap of LP \eqref{knapsacklp} is at least $\frac{d}{2}$.

Since $|P|\le d$ for all $P\in\mathcal{B}$, the Bar-Yehuda--Even algorithm \cite{Bye} 
can be used to find an integral solution to the time-cost tradeoff instance of cost at most $d\cdot \text{LP}$, and can be implemented
to run in polynomial time because for integral vectors $x$ there is a linear-time separation oracle \cite{ByeVlsi}. 
A $d$-approximation can also be obtained by rounding up all $x_v\ge\frac{1}{d}$.

In the following we will improve on this. 
From now on, we assume that we are given a $d$-partition of a hypergraph and an LP solution; 
for time-cost tradeoff instances we get this from Propositions \ref{prop:layers} and \ref{prop:solvelp}.

\section{Rounding fractional vertex covers in \boldmath{$d$}-partite hypergraphs  \label{sec:algorithm}}

In this section, we show how to round a fractional vertex cover in a $d$-partite hypergraph with given $d$-partition. 
Together with the results of the previous section, this yields
an approximation algorithm for time-cost tradeoff instances and will prove Theorem \ref{thm:mainub}.

Our algorithm does not need an explicit list of the edge set of the hypergraph,
which is interesting if $d$ is not constant and there can be exponentially many hyperedges.
The algorithm only requires the vertex set, a $d$-partition, and a feasible solution to the LP (a fractional vertex cover). 
For normalized instances of the time-cost tradeoff problem
such a fractional vertex cover can be obtained as in Proposition \ref{prop:solvelp}, and a $d$-partition by Proposition \ref{prop:layers}.

Our algorithm is based on two previous works for the unweighted $d$-partite hypergraph vertex cover problem.
For rounding a given fractional solution, Lov\'asz \cite{LovaszDiss} obtained a deterministic polynomial-time $(\frac{d}{2}+\epsilon)$-approximation algorithm for any $\epsilon>0$.
Let us quickly sketch his idea.

First, Lov\'asz constructs a family of matrices $A_{d,w}=(a_{ij})_{i=1,\ldots,d,j=0,\ldots,w}$, with the property that: 
\begin{itemize}
\item each row of $A_{d,w}$ is a permutation of $\{0,\ldots,w\}$
\item the sum of each column is at most $\leq \lceil \frac{dw}{2} \rceil$.
\end{itemize}
Now, for a fractional solution $x$ to the $d$-partite hypergraph vertex cover problem, a (large) constant $C$ is chosen, such that $x_v C \in \mathbb{N}$.
The idea is to set $w=\lfloor 2(C-1)/d \rfloor$. Then, for every $j \in \{0,\ldots,w\}$ we may obtain a feasible cover for every $j=0,\ldots,w$ by rounding all $x_v$ for $v \in V_i$ to $1$ if and only if $x_v C > a_{ij}$ (where $V_1,\ldots,V_d$ is the given $d$-partition
of our hypergraph). A simple analysis shows that returning the cheapest such cover is a $\frac{d}{2}\frac{C}{C-1}$ approximation, which converges to $\frac{d}{2}$ for $C \rightarrow \infty$.

Based on this, Aharoni, Holzman and Krivelevich \cite{Aharoni} described a randomized recursive algorithm that works in more general unweighted hypergraphs.
We simplify their algorithm for $d$-partite hypergraphs, which will allow us to obtain
a deterministic polynomial-time algorithm that also works for the weighted problem and always computes a $\frac{d}{2}$-approximation. 
At the end of this section, we will slightly improve on this guarantee in order to compensate for an only approximate LP solution.

We will first describe the algorithm in the even simpler randomized form; then we will derandomize it.
Like Lov\'asz, our algorithm computes a  threshold for each layer to determine whether a variable $x_v$ is rounded up or down.
To compute the random thresholds, and to allow efficient derandomization later, 
we will use a probability distribution with the following properties.

\begin{lem}
\label{lem:prob}
There is a probability distribution that selects
$x \in [0, \frac{2}{9}], y \in [\frac{2}{9}, \frac{4}{9}], z \in [\frac{4}{9},\frac{6}{9}]$, such that
$x+y+z=1$ and $x,y,z$ are uniformly distributed in their respective intervals.
\end{lem}
\begin{pf}
Generate three random numbers in base 3,
$a=0.a_1a_2a_3,\ldots$, $b=0.b_1b_2b_3,\ldots$, $c=0.c_1c_2c_3,\ldots$,
by randomly sampling digits $\{a_i,b_i,c_i\}=\{0,1,2\}$
(that is, we select a random permutation of $\{0,1,2\}$ to be the $i$-th digit of the three numbers).
Let $x'$ be the smallest number, $y'$ the second smallest, and $z'$ the largest number of $a,b,c$.
It is easy to see, that $x' \in [0, \frac{1}{3}]$, $y' \in [\frac{1}{3},\frac{2}{3}]$ and $z' \in [\frac{2}{3}, 1]$.
Also by construction $x'+y'+z'=\frac{3}{2}$. Setting $x=\frac{2}{3}x', y=\frac{2}{3}y', z=\frac{2}{3}z'$ yields the desired result.
\QED
\end{pf}

\begin{figure}[tb]
\centering
\includegraphics[scale=1.0,clip=false]{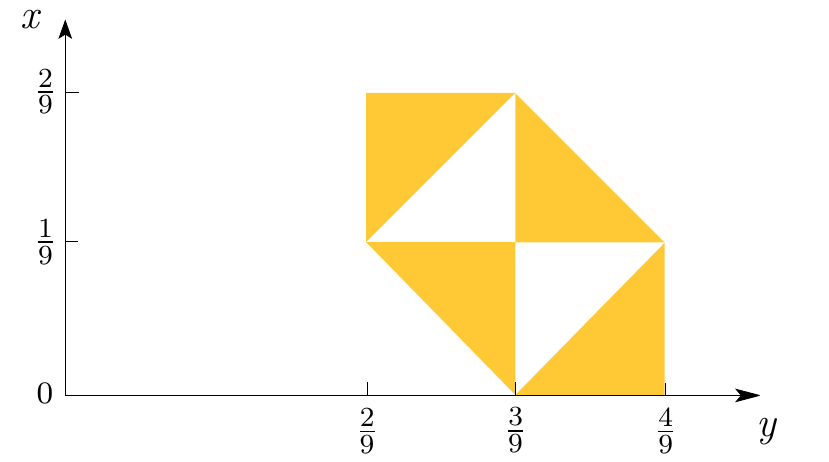}
\caption{Selecting a pair $(x,y)$ by uniformly sampling a point in the yellow area
gives an example of how to choose random numbers $x,y$ (and $z=1-x-y$) as in Lemma \ref{lem:prob}.}
\label{fig:randomness}
\end{figure}

We remark that for implementing an algorithm that samples from this distribution, a different construction is more suitable.
For example, one may start by selecting $x \in [0, \frac{2}{9}]$ randomly, and then use a case distinction as illustrated in Figure \ref{fig:randomness}
to select $y$ randomly in a suitable subset of $[\frac{2}{9}, \frac{4}{9}]$. Finally, we may set $z=1-x-y$. It is easy to verify that this also achieves the claimed properties.

For our proof we will need to slightly generalize this distribution to an arbitrary number of elements.
\begin{lem}
\label{lem:prob_extended}
For any $d \geq 2$, there is a probability distribution that selects $a_1,\ldots,a_d$, such that
$\sum_{i=1}^d a_i=1$ and $a_i$ is uniformly distributed in $[\frac{2(i-1)}{d^2}, \frac{2i}{d^2}]$.
For any $i,j$ such that $|i-j| \ge 3$, the random variables corresponding to $a_i$ and $a_j$ are independent.
\end{lem}
\begin{pf}
For $d=2$ we can just choose $a_1$ uniformly in $[0,\frac{1}{2}]$ and set $a_2=1-a_1$.
The case $d=3$ follows from Lemma \ref{lem:prob}. In general, note that the sum of the
expectations of the $a_i$ is $\sum_{i=1}^d \frac{2i-1}{d^2}= 1$.
Hence we can partition $1,\ldots,d$ into groups of two or three and apply the above
with appropriate scaling and shifting.

More precisely, if $d$ is odd, we choose $x,y,z$ according to Lemma \ref{lem:prob} and set
$a_1=\frac{9x}{d^2}$, $a_2=\frac{9y}{d^2}$, and $a_3=\frac{9z}{d^2}$.
Then the remaining number of indices is even, and we group them into pairs;
for indices $i$ and $i+1$ we choose $a_i$ uniformly in $[\frac{2(i-1)}{d^2}, \frac{2i}{d^2}]$ and
set $a_{i+1}:=\frac{4i}{d^2}-a_i$.
\QED
\end{pf}

\begin{thm}\label{thm:k2approx_rnd}
Let $x$ be a fractional vertex cover in a $d$-partite hypergraph with given $d$-partition.
There is a randomized linear-time algorithm that computes an integral solution $\bar x$ of expected cost
$\mathbb{E}[\sum_{v\in V}c_v\cdot \bar x_v] \le \frac{d}{2} \sum_{v\in V}c_v\cdot x_v$.
\end{thm}

\begin{figure*}[tb!]
\centering
\includegraphics[scale=0.6,clip=false]{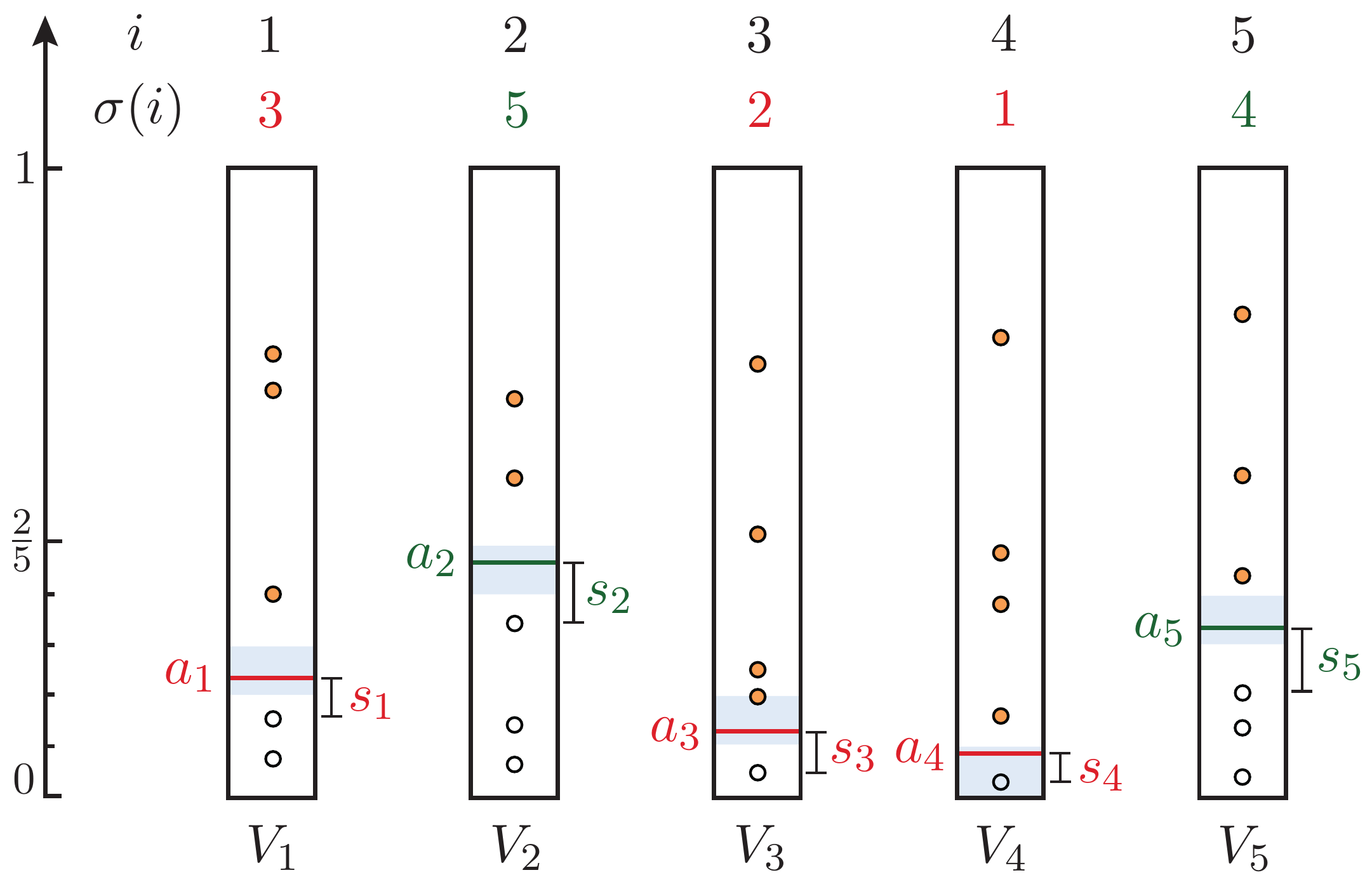}
\caption{
A sketch of thresholds $a_1,\ldots,a_5$ chosen by our randomized algorithm in Theorem~\ref{thm:k2approx_rnd} for the case $d=5$.
The circles represent vertices in the hypergraph, drawn by their position in the partition and the value of their corresponding variable in the LP.
Suppose the permutation $(\sigma(1),\ldots,\sigma(5))=(3,5,2,1,4)$ is chosen.
Then the thresholds $a_i$ are randomly chosen in the light blue intervals $\bigl[ \frac{2(\sigma(i)-1)}{d^2}, \frac{2\sigma(i)}{d^2} \bigr]$; 
moreover, the thresholds $a_1,a_3,a_4$ are chosen independently of the thresholds $a_2,a_5$, as indicated by their color.
 The points above the thresholds are filled; these variables are rounded up to $1$, while the empty circles represent variables that are
 rounded down to $0$.
Finally, the figure also shows ``slack'' values $s_1,\ldots,s_5$, telling how much each threshold could be lowered 
without changing the solution returned by our algorithm. 
These will play a key role to improve the approximation guarantee in Theorem \ref{thm:betterthank2approx}.
}
\label{fig:thresholds}
\end{figure*}

\begin{pf}
Let $V_1,\ldots,V_d$ be the given $d$-partition of our hypergraph $(V,\mathcal{B})$, so $|P \cap V_i|\leq 1$ for all $i=1,\ldots,d$
and every hyperedge $P\in\mathcal{B}$.
We write $l(v)=i$ if $v\in V_i$ and call $V_i$ a \textit{layer} of the given hypergraph.

Now consider the following randomized algorithm, which is also illustrated in Figure \ref{fig:thresholds}:
Choose a random permutation $\sigma:\{1,\ldots,d\}\to\{1,\ldots,d\}$ and choose random numbers
$a_i$ uniformly distributed in $\bigl[ \frac{2(\sigma(i)-1)}{d^2}, \frac{2\sigma(i)}{d^2} \bigr]$ for $i=1,\ldots,d$
such that $\sum_{i=1}^d a_i=1$, as constructed in Lemma \ref{lem:prob_extended}.
Then, for all $v\in V$, set $\bar x_v:=1$ if $x_v\ge a_{l(v)}$ and $\bar x_v:=0$ if $x_v<a_{l(v)}$.

To show that $\bar x$ is a feasible solution, observe that any hyperedge $P\in\mathcal{B}$
has $\sum_{v\in P} x_v\ge 1 = \sum_{i=1}^d a_i \ge \sum_{v\in P} a_{l(v)}$
and hence $x_v\ge a_{l(v)}$ for some $v\in P$.

It is also easy to see that the probability that $\bar x_v$ is set to 1 is exactly $\min\{1,\frac{d}{2}x_v\}$.
Indeed, if $x_v\ge\frac{2}{d}$, we surely set $\bar x_v=1$.
Otherwise, $x_v\in \bigl[ \frac{2(j-1)}{d^2}, \frac{2j}{d^2} \bigr]$ for some $j\in\{1,\ldots,d\}$;
then we set $\bar x_v=1$ if and only if $\sigma(l(v))<j$ or ($\sigma(l(v))=j$ and $a_{l(v)}\le x_v$),
which happens with probability
$\frac{j-1}{d} + \frac{1}{d} (x_v-\frac{2(j-1)}{d^2})\frac{d^2}{2}=\frac{d}{2}x_v$.
Hence the expected cost
$\mathbb{E}[\sum_{v\in V}c_v\cdot \bar x_v]$ is at most $\frac{d}{2} \sum_{v\in V}c_v\cdot x_v$. \QED
\end{pf}

Now we derandomize this algorithm and show how to implement it in polynomial time.

\begin{thm}\label{thm:k2approx}
Let $x$ be a fractional vertex cover in a $d$-partite hypergraph with given $d$-partition.
There is a deterministic algorithm that computes an integral solution $\bar x$ of cost
$\sum_{v\in V}c_v\cdot \bar x_v \le \frac{d}{2} \sum_{v\in V}c_v\cdot x_v$
in time $\mathcal{O}(n^3)$.
\end{thm}
\begin{pf}
For a fixed value $\sigma(i)=j$ and a random choice of
$a_i \in \bigl[ \frac{2(j-1)}{d^2}, \frac{2j}{d^2} \bigr]$ we have the expected cost
$$\mathbb{E}\left[ \bar x_v \mid \sigma(i)=j \right] = \left\{\begin{array}{ll}
        0, & \text{if }  x_v < \frac{2(j-1)}{d^2}\\
        x_v-\frac{2(j-1)}{d^2} \cdot \frac{d^2}{2}, & \text{if }  x_v \in \bigl[ \frac{2(j-1)}{d^2}, \frac{2j}{d^2} \bigr]\\
        1 & \text{if }  x_v > \frac{2j}{d^2}
        \end{array}\right.$$

Let $\rho(i,j):=\sum_{v \in V_i}c_v\cdot\mathbb{E}\left[\bar x_v \mid \sigma(i)=j \right]$ be the total expected cost of layer $i$
if we assign $\sigma(i)=j$ in the random permutation.
We compute a permutation $\sigma$ that minimizes the total expected cost
$\sum_{i=1}^d \rho(i,\sigma(i))$; this is a minimum-cost perfect matching problem in a complete bipartite graph
with $d+d$ vertices.
Hence this step can be implemented with a running time of $\mathcal{O}(d^3)$ \cite{EdmondsKarp,Tomizawa}.

Therefore, we may now assume that the permutation $\sigma$ is fixed.
The probability distribution described in Lemma~\ref{lem:prob_extended}
chooses the values $a_i$ for $i \in \{1,\ldots,d\}$ independently for groups of two or three layers,
with fixed sum $S_I:=\sum_{i\in I}a_i$ for each such group $I$.
Setting $a'_i=\max\{x_v:v\in V_i,\,x_v < a_i\}$, we see that the result in group $I$ 
depends only on the numbers $a'_i$ ($i\in I$)
and that there are less than $n^3$ possibilities.
Among all choices of the $a'_i$ ($i\in I$) with $\sum_{i\in I}a'_i<S_I$, we can thus choose an optimum one
(with minimum $\sum_{i\in I}\sum_{v\in V_i: x_v>a'_i}c_v$) in $\mathcal{O}(n^3)$ time.
\QED
\end{pf}

It is easy to improve the running time in Theorem \ref{thm:k2approx} to $\mathcal{O}(d^3+n^2/d^2)$, 
but this is not important since already for time-cost tradeoff instances solving the LP
dominates the overall running time of our approximation algorithm.

Since the vertex cover LP can be solved only approximately (Proposition~\ref{prop:solvelp}), 
this would only yield an approximation ratio of $\frac{d}{2}+\epsilon$ for the time-cost tradeoff problem (unless $d$ is fixed). 
In order to obtain a true $\frac{d}{2}$-approximation algorithm (and thus prove  Theorem~\ref{thm:mainub}),
we need a slightly stronger bound, which we derive next. 
Again, we first describe an improved randomized algorithm and then derandomize it.

\begin{thm}\label{thm:betterthank2approx}
Let $d\ge 4$.
Let $x$ be a fractional vertex cover in a $d$-partite hypergraph with given $d$-partition.
There is a randomized linear-time algorithm that computes an integral solution $\bar x$ of expected cost
$\sum_{v\in V}c_v\cdot \bar x_v \le (\frac{d}{2}-\frac{d}{64n}) \sum_{v\in V}c_v\cdot x_v$.
\end{thm}

\begin{pf}
First we choose the permutation $\sigma$ and thresholds $a_1,\ldots,a_d$ with $\sum_{i=1}^d a_i=1$ 
randomly as above such that the thresholds are independent except within groups of two or three.
For $i\in\{1,\ldots,d\}$ denote the \emph{slack} of level $i$ by 
$s_i:=\min\{\frac{1}{d}, a_i,\, a_i - \max\{x_v: v\in V_i,\, x_v<a_i\}\}$.
The slack  is always non-negative.
Lowering the threshold $a_i$ by less than $s_i$ would yield the same solution $\bar x$.
The reason for cutting off the slack at $\frac{1}{d}$ will become clear only below.

Next we randomly select one level $\lambda\in\{1,\ldots,d\}$.
Let $\Lambda$ be the corresponding group (cf.\ Lemma~\ref{lem:prob_extended}), i.e., 
$\lambda\in\Lambda\subseteq\{1,\ldots,d\}$, $|\Lambda|\le 3$, and 
$a_i$ is independent of $a_{\lambda}$ whenever $i\notin\Lambda$.
Now raise the threshold $a_{\lambda}$
to $a'_{\lambda}=a_{\lambda}+\sum_{i\notin\Lambda} s_i$.
Set $a'_i=a_i$ for $i\in\{1,\ldots,d\}\setminus\{\lambda\}$.

As before, for all $v\in V$, set $\bar x_v:=1$ if $x_v\ge a'_{l(v)}$ and $\bar x_v:=0$ if $x_v<a'_{l(v)}$.
We first observe that $\bar x$ is feasible.
Indeed, if there were any hyperedge $P\in\mathcal{B}$ with $x_v<a'_{l(v)}$ for all $v\in P$, we would get 
$1 \le \sum_{v\in P} x_v < \sum_{v\in P:l(v)\notin\Lambda} (a_{l(v)} - s_{l(v)}) + \sum_{v\in P:l(v)\in\Lambda} a'_{\lambda}
\le \sum_{i\notin\Lambda} (a_i - s_i) +\sum_{i\in\Lambda\setminus\{\lambda\}} a_i + a'_{\lambda} = \sum_{i=1}^d a_i = 1$, a contradiction. 

We now bound the expected cost of $\bar x$. 
Let $v\in V$.
With probability $\frac{d-1}{d}$ we have $l(v)\not=\lambda$ and, conditioned on this, an expectation 
$\mathbb{E}\left[\bar x_v \mid \lambda\not=l(v)\right] = \frac{d}{2}\min\{x_v,\frac{2}{d}\}\le \frac{d}{2}x_v$ as before.
Now we condition on $l(v)=\lambda$ and in addition, 
for any $S$ with $0\le S\le \frac{d-2}{d}$, on $\sum_{i\notin\Lambda} s_i=S$; note that $a_{\lambda}$ is independent of $S$.
The probability that $\bar x_v$ is set to 1 is 
$\frac{d}{2}\max\{0,\,\min\{x_v-S,\frac{2}{d}\}\} \le \frac{x_v}{\frac{2}{d}+S} \le \frac{d}{2}(1-S)x_v$ in this case.
In the last inequality we used $S\le \frac{d-2}{d}$, and this was the reason to cut off the slacks.
In total we have for all $v\in V$:
\begin{align*}
\mathbb{E}\left[\bar x_v\right] \ = \ & \frac{d-1}{d} \cdot \mathbb{E}[\bar x_v \mid \lambda\not=l(v)] \\
& + \frac{1}{d} \cdot \int_{0}^{\frac{d-2}{d}} \mathbb{P}\left[\sum_{i\notin\Lambda} s_i=S \mid \lambda=l(v)\right] \cdot 
\mathbb{E}\left[\bar x_v \mid \lambda=l(v),\, \sum_{i\notin\Lambda} s_i=S \right] \text{ d}S \\
\ \le \ & \frac{d-1}{d} \cdot \frac{d}{2}x_v 
+ \frac{1}{d} \cdot \int_{0}^{\frac{d-2}{d}} \mathbb{P}\left[\sum_{i\notin\Lambda} s_i=S \mid \lambda=l(v)\right] \cdot
\frac{d}{2}(1-S)x_v \text{ d}S \\
\ \le \ & \frac{d}{2} \left( 1 - \frac{1}{d}
\int_{0}^{\frac{d-2}{d}} \mathbb{P}\left[\sum_{i\notin\Lambda} s_i=S \mid \lambda=l(v)\right] \cdot S \text{ d}S \right) \cdot x_v \\
\ = \ & \frac{d}{2} \left(1 - \frac{1}{d} \cdot \mathbb{E}\left[S \mid \lambda=l(v)\right] \right) x_v.
\end{align*}
Let $\Lambda[v]$ be the the set $\Lambda$ in the event $\lambda=l(v)$. We estimate 
$$\mathbb{E}\left[S\mid \lambda=l(v)\right] = \sum_{i\notin\Lambda(v)} \mathbb{E}\left[s_i\right] 
\ge \sum_{i\notin\Lambda(v)} \frac{1}{d(n_i+1)} 
\ge \frac{(d-3)^2}{d(n+d)} \ge \frac{d}{32n}.$$
Here $n_i=|V_i|$, and the first inequality holds because $\mathbb{E}\left[s_i\right]$ is maximal if 
$\{x_v:v\in V_i\}=\{\frac{2j}{d(n_i+1)}: j=1,\ldots,n_i\}$.
We conclude $\mathbb{E}\left[\sum_{v\in V}c_v\cdot \bar x_v\right] \le ( \frac{d}{2} - \frac{d}{64n} ) \sum_{v\in V}c_v\cdot x_v$.
\QED
\end{pf}

Let us now derandomize this algorithm. This is easier than before because we can afford to lose a little again.

\begin{thm}\label{thm:betterthank2approx-deterministic}
Let $d\ge 4$.
Let $x$ be a fractional vertex cover in a $d$-partite hypergraph with given $d$-partition.
There is a deterministic algorithm that computes an integral solution $\bar x$ of cost
$\sum_{v\in V}c_v\cdot \bar x_v \le (\frac{d}{2}-\frac{d}{128n}) \sum_{v\in V}c_v\cdot x_v$
in time $\mathcal{O}(n^3)$.
\end{thm}

\begin{pf}
Let again $\text{LP}=\sum_{v\in V}c_v\cdot x_v$ denote the LP value.
We first round down the costs to integer multiples of $\frac{d\,\text{LP}}{128 n^2}$ by setting
$c'_v:= \lfloor \frac{128 n^2 c_v}{d\,\text{LP}} \rfloor\frac{d\,\text{LP}}{128 n^2}$ for $v\in V$. 
Then we compute the best possible choice of threshold values 
$a_i$ for $i \in \{1, \ldots, d\}$ such that $\sum_{j=1}^d a_j \leq 1$ and 
$\sum_{j=1}^d \sum_{v \in V_j, x_v \geq a_j}c'_v$ is minimized. 
This is a simple dynamic program (like for the knapsack problem) that runs in 
$\mathcal{O}(n^3)$ time.
By Theorem~\ref{thm:betterthank2approx} there is such a solution with cost
$\sum_{j=1}^d \sum_{v \in V_j, x_v \geq a_j}c'_v \le \sum_{j=1}^d \sum_{v \in V_j, x_v \geq a_j}c_v 
\le ( \frac{d}{2} - \frac{d}{64n} ) \text{LP}$.
Hence the solution that we find costs
$\sum_{j=1}^d \sum_{v \in V_j, x_v \geq a_j}c_v 
< \sum_{j=1}^d \sum_{v \in V_j, x_v \geq a_j}c'_v  + n\frac{d\,\text{LP}}{n^2}
\le ( \frac{d}{2} - \frac{d}{64n} ) \text{LP}  + \frac{d\,\text{LP}}{128n}$
as required.
\QED
\end{pf}

As explained above, together with Propositions \ref{prop:layers} and \ref{prop:solvelp}
(with $\epsilon=\frac{1}{128n}$), Theorem \ref{thm:betterthank2approx-deterministic} 
implies Theorem~\ref{thm:mainub}.

\section{Inapproximability \label{sec:inapprox}}

Guruswami, Sachdeva and Saket \cite{InapproxHypGraph} proved that approximating the vertex cover problem in $d$-partite hypergraphs
with a better ratio than $\frac{d}{2}$ is NP-hard under the Unique Games Conjecture.
We show that even for the special case of time-cost tradeoff instances, the problem is hard to approximate by a factor of $\frac{d+2}{4}$.

Note that this is really a special case: for example the 3-partite hypergraph with vertex set $\{1,2,3,4,5,6\}$ and hyperedges $\{1,4,6\}, \{2,3,6\}$, and $\{2,4,5\}$ does not result from a time-cost tradeoff instance of depth 3 with our construction.\footnote{
Assume we have an instance of the time-cost tradeoff instance on the same vertex set 
such that the minimal chains requiring speedup correspond to precisely these hyperedges.
For $v \in \{1,2,3,4,5,6\}$ let $l(v) \in \{1,2,3\}$ denote the layer of job $v$.
Note that jobs 2, 4, and 6 must belong to distinct layers.
Since the pairs $(1,2)$, $(3,4)$, and $(5,6)$ are symmetric 
(any permutation of these pairs yields an automorpishm of $(V,\mathcal{B})$),  
we may assume without loss of generality $l(2)<l(4)<l(6)$.
Then $l(1)=l(2)=1$, $l(3)=l(4)=2$, and $l(5)=l(6)=3$.

But then $\{1,4,5\}$ and $\{2,4,6\}$ are chains, and their total delay equals the total delay
of the chains $\{1,4,6\}$ and $\{2,4,5\}$. 
Since the latter two chains exceed the deadline, 
at least one of the former two chains must also exceed the deadline.
This is a contradiction since neither $\{1,4,5\}$ nor $\{2,4,6\}$ nor any subset is in $\mathcal{B}$.
}

Let us briefly sketch a technique of \cite{InapproxHypGraph} and explain why it does not serve our purpose.
Let $d\geq k \geq 2$ be integers. One can reduce the vertex cover problem in $k$-uniform hypergraphs, 
i.e., for hypergraphs $H=(U,F)$
such that $|e|=k$ for all $e \in F$, to the $d$-partite case. The idea is to take $d$ disjoint copies of the vertex
set $U$ as the vertex set of a new hypergraph $G$. For every hyperedge $e \in F$, the hypergraph $G$ contains all hyperedges $e'$ that contain exactly 
one copy of every vertex in $e$ and at most one vertex of any of the $d$ copies of $U$.
Clearly this hypergraph $G$ is $d$-partite. 
It is easy to see that any optimal solution in $G$ must contain either no or at least $d-k+1$ of the copies 
of a vertex and there is always a vertex cover of size $d\cdot\text{OPT}$, where
$\text{OPT}$ denotes the size of an optimum vertex cover in $H$.
By a result of Khot and Regev \cite{Khot}, the vertex cover problem in $k$-uniform hypergraphs is NP-hard to approximate with a factor of $k-\epsilon$ under the Unique Games Conjecture.
Therefore, for any $d \geq 4$, by letting $k=\lceil \frac{d+1}{2} \rceil$, we obtain a $d$-partite hypergraph vertex-cover instance. From this one can conclude that these instances do not admit a $\frac{d}{4}$-approximation algorithm (assuming the Unique Games Conjecture and P$\neq$NP). However, as indicated above, these instances are more general than those resulting from time-cost tradeoff instances of depth $d$. Nevertheless we will use some of these ideas below.

In this and the next section, we will show Theorem \ref{thm:mainlb}, which is our main inapproximability result.
Insetad of starting from $k$-uniform hypergraphs,
we devise a reduction from the vertex deletion problem in acyclic digraphs, which
Svensson \cite{Svensson} called DVD.\footnote{An undirected version of this problem has been called
\emph{$k$-path vertex cover}\cite{Bresaretal} or \emph{vertex cover $P_k$} \cite{TuZhou}.}
Let $k$ be a positive integer; then $\textnormal{DVD}(k)$ is defined as follows:
given an acyclic digraph, compute a minimum-cardinality set of vertices whose deletion destroys all paths with $k$ vertices.
This problem is easily seen to admit a $k$-approximation algorithm:

\begin{lem}
\label{lem:bye_vdel}
For all $k\ge 1$, $\textnormal{DVD}(k)$ admits a $k$-approximation algorithm.
\end{lem}

\begin{pf}
Find a maximal set of vertex-disjoint paths, each with $k$ vertices, and take the set of all their vertices.
\QED
\end{pf}

Svensson proved that anything better than this simple approximation algorithm would solve the unique games problem:

\begin{thm}[{\cite{Svensson}}]
\label{thm:khardness}
Let $k \in \mathbb{N}$ with $k \geq 2$ and $\rho<k$ be constants.
Let $\text{OPT}$ denote the size of an optimum solution for a given $\textnormal{DVD}(k)$ instance.
Assuming the Unique Games Conjecture it is NP-hard to compute a number $l \in \mathbb{R}_+$ such that
$l \leq \text{OPT} \leq \rho l$.
\end{thm}

This is the starting point of our proof.
Svensson \cite{Svensson} already observed that
$\textnormal{DVD}(k)$ can be regarded as a special case of the time-cost tradeoff problem.
Note that this does not imply Theorem \ref{thm:mainlb} because
the hard instances of $\textnormal{DVD}(k)$ constructed in the proof of Theorem \ref{thm:khardness}
have unbounded depth even for fixed $k$.
(Recall that the \emph{depth} of an acyclic digraph is the number of vertices in a longest path.)
The following is a variant (and slight strengthening) of Svensson's observation.

\begin{figure*}[tb]
\centering
\includegraphics[scale=1.45,clip=false]{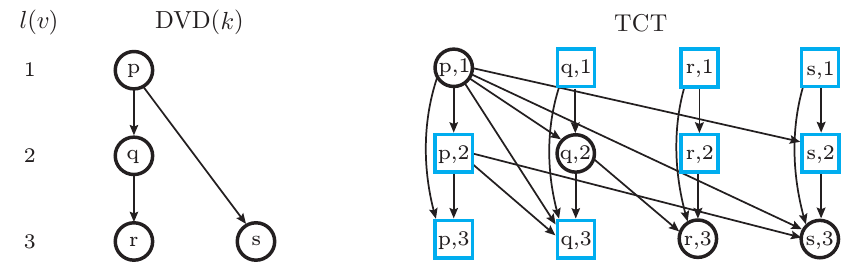}
\caption{The transformation of Lemma \ref{lem:modeldvdastct}.
An instance of $\textnormal{DVD}(k)$ is transformed into an equivalent instance of the time-cost tradeoff problem.
Jobs with fixed execution time are depicted as blue squares.}
\label{fig:tct}
\end{figure*}

\begin{lem}
\label{lem:modeldvdastct}
Any instance of $\textnormal{DVD}(k)$ (for any $k$) can be transformed in linear time
to an equivalent instance of the time-cost tradeoff problem, with the same depth and the same optimum value.
\end{lem}

\begin{pf}
Let $G=(V,E)$ be an instance of $\textnormal{DVD}(k)$, an acyclic digraph, say of depth $d$.
Let $l(v)\in\{1,\ldots,d\}$ for $v\in V$ such that $l(v)<l(w)$ for all $(v,w)\in E$.
Let $J:=\{(v,i): v\in V,\, i\in \{1,\ldots,d\}\}$ be the set of jobs of our time-cost tradeoff instance.
Job $(v,i)$ must precede job $(w,j)$ if ($v=w$ and $i<j$) or ($(v,w)\in E$ and $l(v)\le i <j$).
Let $\prec$ be the transitive closure of these precedence constraints.
For $v\in V$, the job $(v,l(v))$ is called \emph{variable} and has a fast execution time 0
at cost 1 and a slow execution time $d+1$ at cost 0.
All other jobs are \emph{fixed}; they have a fixed execution time $d$ at cost 0.
The deadline is $d^2+k-1$. A sketch of this construction is given in Figure \ref{fig:tct}.

We claim that any set of variable jobs whose acceleration constitutes a feasible solution
of this time-cost tradeoff instance corresponds
to a set of vertices whose deletion destroys all paths in $G$ with $k$ vertices, and vice versa.
Indeed, the total delay of a chain in the time-cost tradeoff instance is at most $(d-1)(d+1)$
unless the chain contains a job in each level and contains no variable job that is accelerated,
in which case the total delay is $d^2 + j$, where $j$ is the number of variable jobs in the chain.
These chains with total delay $d^2+j$ correspond to the paths with $j$ vertices in $G$.
\QED
\end{pf}

Therefore a hardness result for $\textnormal{DVD}(k)$ for bounded depth instances transfers to
a hardness result for the time-cost tradeoff problem with bounded depth.
We will show the following strengthening of Theorem \ref{thm:khardness}:

\begin{thm}\label{thm:dvdhardnessforboundeddepth}
Let $k,d \in \mathbb{N}$ with $2\leq k \leq d$ and $\rho<\frac{k(d+1-k)}{d}$ be constants.
Let $\text{OPT}$ denote the size of an optimum solution for a given $\textnormal{DVD}(k)$ instance.
Assuming the Unique Games Conjecture it is NP-hard to compute a number $l \in \mathbb{R}_+$ such that
$l \leq \text{OPT} \leq \rho l$.
\end{thm}

It is easy to see that Theorem \ref{thm:dvdhardnessforboundeddepth} and Lemma \ref{lem:modeldvdastct}
imply Theorem \ref{thm:mainlb}. Indeed, let $d\in\mathbb{N}$ with $d\ge 2$ and $\rho<\frac{d+2}{4}$,
and suppose that a $\rho$-approximation algorithm $\mathcal{A}$ exists for time-cost tradeoff instances of depth $d$.
Let $k:=\lceil\frac{d+1}{2}\rceil$ and consider an instance of $\textnormal{DVD}(k)$ with depth $d$.
Transform this instance to an equivalent time-cost tradeoff instance by Lemma \ref{lem:modeldvdastct}
and apply algorithm $\mathcal{A}$.
This constitutes a $\rho$-approximation algorithm for $\textnormal{DVD}(k)$ with depth $d$.
Since $\rho<\frac{d+2}{4}\le\frac{k(d+1-k)}{d}$, Theorem \ref{thm:dvdhardnessforboundeddepth} then implies
that the Unique Games Conjecture is false or $\textnormal{P}=\textnormal{NP}$.\footnote{In fact,
this proof shows that the threshold in Theorem \ref{thm:mainlb} can be taken $\frac{1}{4d}$ larger for odd $d$;
e.g., there is no $\rho$-approximation algorithm for $\rho<\frac{4}{3}$ for $d=3$.}

It remains to prove Theorem \ref{thm:dvdhardnessforboundeddepth}, which will be the subject of the next section.

\section{Reducing vertex deletion to constant depth \label{sec:depthreduction}}

In this section we prove Theorem \ref{thm:dvdhardnessforboundeddepth}.
The idea is to reduce the depth of a digraph by transforming it to another digraph with small depth but
related vertex deletion number.
Let $k,d\in\mathbb{N}$ with $2\leq k\leq d$, and let $G$ be a digraph.
We construct an acyclic digraph $G^d$ of depth at most $d$ by taking the tensor product with the acyclic tournament on $d$ vertices:
$G^d=(V^d,E^d)$, where
$V^d=V\times\{1,\ldots,d\}$ and $E^d=\{((v,i),(w,j)) : (v,w)\in E \text{ and } i<j\}$.
It is obvious that $G^d$ has depth $d$. An example of this construction is depicted in Figure \ref{fig:tensor}.
Here is our key lemma:

\begin{figure*}[tb]
\centering
\includegraphics[scale=0.98,clip=false]{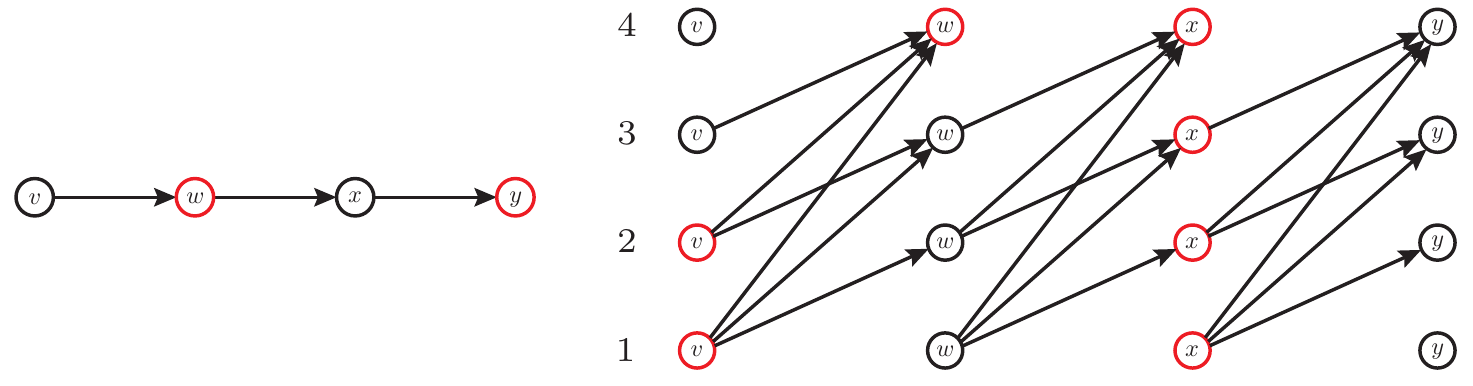}
\caption{A directed path $P_4$ and the graph tensor product with the acyclic tournament on $4$ vertices. The colored vertices show a solution to the vertex deletion problems with $k=2$.}
\label{fig:tensor}
\end{figure*}

\begin{lem}
Let $G$ be an acyclic directed graph and $k,d\in\mathbb{N}$ with $2\le k\le d$.
If we denote by $\text{OPT}(G,k)$ the minimum number of vertices of $G$ hitting all paths with $k$ vertices,
then
\begin{equation}\label{eq:optDVDtensor}
(d+1-k)\cdot \text{OPT}(G,k) \ \leq \ \text{OPT}(G^d,k) \ \leq \ d \cdot\text{OPT}(G,k).
\end{equation}
\label{lem:tensorlemma}
\end{lem}

Lemma \ref{lem:tensorlemma}, together with Theorem \ref{thm:khardness}, immediately implies
Theorem \ref{thm:dvdhardnessforboundeddepth}:
assuming a $\rho$-approximation algorithm for $\textnormal{DVD}(k)$ instances with depth $d$,
with $\rho<\frac{k(d+1-k)}{d}$, we can compute $\text{OPT}(G,k)$ up to a factor less than $k$ for
any digraph $G$. By Theorem \ref{thm:khardness}, this would contradict the Unique Games Conjecture
or $\text{P}\not=\text{NP}$.

Before we prove Lemma \ref{lem:tensorlemma}, let us give two examples that show
that the bounds in \eqref{eq:optDVDtensor} are sharp for all $d$ and $k$,
for infinitely many acyclic digraphs.

For the lower bound, consider the acyclic tournament $D_n$ on the vertices $1,\ldots,n$.
Obviously, $\text{OPT}(D_n,k)=n-k+1$. Moreover,
$\text{OPT}(D_n^d,k) \le (d+1-k)(n-k+1)$ because
$\{(i,j): i=1,\ldots,n-k+1,\, j=1,\ldots, d+1-k\}$ is a feasible solution for $\text{DVD}(D_n^d,k)$.

\begin{figure*}[tb]
\centering
\includegraphics[scale=1.2,clip=false]{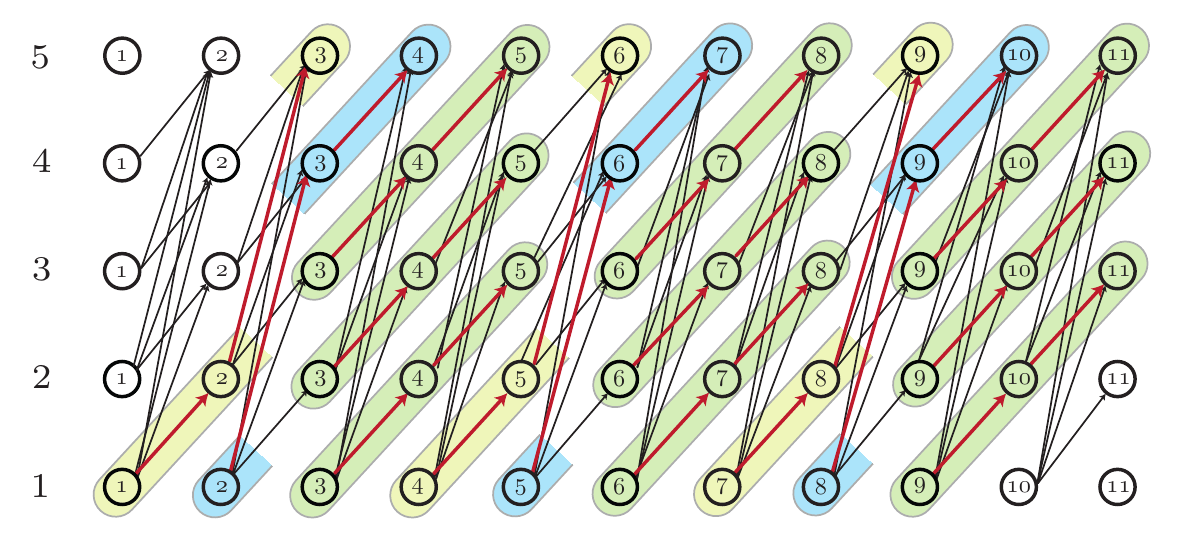}
\caption{Construction of $r d$ vertex-disjoint paths, each with $k$ vertices,
in $P_{(r+1)k-1}^d$ for $r=3$, $d=5$, and $k=3$.
The edge sets corresponding to paths are highlighted in red.}
\label{fig:ex_packing}
\end{figure*}

For the upper bound, consider the directed path $P_n$ on the vertices $1,\ldots,n$, where
$n=(r+1)k-1$ for some $r \in \mathbb{N}$. Obviously $\text{OPT}(P_n,k)=r$ because
$\{k,2k,\ldots,rk\}$ is a feasible solution. To show $\text{OPT}(P_n^d,k) \ge rd$,
we find $rd$ vertex-disjoint paths in $P_n^d$, each with $k$ vertices:
for $i=1,\ldots,r$ and $j=1,\ldots,d$,
the vertex set of the $(di-d+j)$-th path arises from $\{(ki,j),(ki+1,j+1),\ldots,(ki+k-1,j+k-1)\}$
by replacing $(s,d+t)$ by $(s-k+t,t)$ for all $s,t\ge 1$.
See Figure \ref{fig:ex_packing}.

We remark that the left inequality in \eqref{eq:optDVDtensor} holds also for general (not necessarily acyclic) digraphs.
However, for general digraphs it may be that $\text{OPT}(G^d,k)>d \cdot\text{OPT}(G,k)$.

Finally, we prove Lemma \ref{lem:tensorlemma}.

\begin{pf}{(Lemma \ref{lem:tensorlemma})}
Let $G$ be an acyclic digraph.
The upper bound of \eqref{eq:optDVDtensor} is trivial:
for any set $W\subseteq V$ that hits all $k$-vertex paths in $G$
we can take $X:=W\times\{1,\ldots,d\}$ to obtain a solution to the $\textnormal{DVD}(k)$ instance $G^d$.

To show the lower bound,
we fix a minimal solution $X$ to the $\textnormal{DVD}(k)$ instance $G^d$.
Let $Q$ be a path in $G^d$ with at most $k$ vertices. We write $\text{start}(Q)=i$
if $Q$ begins in a vertex $(v,i)$.
We define $\mathcal{Q}$ as the set of paths in $G^d$ with exactly $k$ vertices. 
For $Q\in\mathcal{Q}$ let $\text{lasthit}(Q)$ denote the last vertex of $Q$ that belongs to $X$.
For $x\in X$ we define
$$\varphi(x) := \max\{\text{start}(Q): Q\in\mathcal{Q},\, \text{lasthit}(Q) = x\}.$$
Note that this is well-defined due to the minimality of $X$, and $1\le \varphi(x) \le d+1-k$ for all $x\in X$.

We will show that for $j=1,\ldots,d+1-k$,
$$S_j := \{v\in V: (v,i)\in X \text{ and } \varphi((v,i))=j \text{ for some } i\in\{1,\ldots,d\}\}$$
hits all $k$-vertex paths in $G$.
This shows the lower bound in \eqref{eq:optDVDtensor} because then
$\text{OPT}(G,k)\le \min_{j=1}^{d+1-k} |S_j| \le \frac{|X|}{d+1-k}$.

Let $P$ be a path in $G$ with $k$ vertices $v_1,\ldots,v_k$ in this order.
Consider $d$ ``diagonal'' copies $D_1,\ldots,D_d$ of (suffixes of) $P$ in $G^d$:
the path $D_i$ consists of the vertices $(v_s,s+i-k),\ldots,(v_k,i)$, where $s=\max\{1,k+1-i\}$.
Note that the paths $D_1,\ldots,D_{k-1}$ have fewer than $k$ vertices.

We show that for each $j=1,\ldots,d+1-k$,
at least one of these diagonal paths contains a vertex $x\in X$ with $\varphi(x)=j$.
This implies that $S_j\cap P\not=\emptyset$ and concludes the proof.

First,
$D_d$ contains a vertex in $x\in X$ with $\varphi(x)=d+1-k$, namely $\text{lasthit}(D_d)$.
Now we show for $i=1,\ldots,d-1$ and $j=1,\ldots,d-k$:

\medskip\noindent\textbf{Claim:}
If $D_{i+1}$ contains a vertex $x\in X$ with $\varphi(x)=j+1$, then
$D_i$ contains a vertex $x'\in X$ with $\varphi(x')\geq j$.
\medskip

This Claim implies the theorem because $D_1$ consists of a single vertex $(v_k,1)$, and if it belongs to $X$,
then $\varphi((v_k,1))=1$.

To prove the Claim (see Figure \ref{fig:proof} for an illustration), let
$x=(v_h,l(x))\in X\cap D_{i+1}$ and $\varphi(x)\geq j+1$,
and let $x$ be the last such vertex on $D_{i+1}$.
We have $\varphi(x)\geq \text{start}(D_{i+1})$ for otherwise we have $\text{start}(D_{i+1})>1$, so $D_{i+1}$ contains $k$ vertices and we should
have chosen $x=\text{lasthit}(D_{i+1})$; note that $\varphi(\text{lasthit}(D_{i+1}))\geq\text{start}(D_{i+1})$.

Let $Q\in\mathcal{Q}$ be a path attaining the maximum in the definition of $\varphi(x)$.
So $\text{start}(Q)=\varphi(x)$ and $\text{lasthit}(Q)=x$.
Suppose $x$ is the $p$-th vertex of $Q$; note that
\begin{equation}\label{eq:boundposition}
p \ \le \ 1+l(x)-\varphi(x)
\end{equation}
because $Q$ starts on level $\varphi(x)$, rises at least one level with every vertex,
and reaches level $l(x)$ at its $p$-th vertex.

\begin{figure*}[tb]
\centering
\includegraphics[scale=1.0,clip=false]{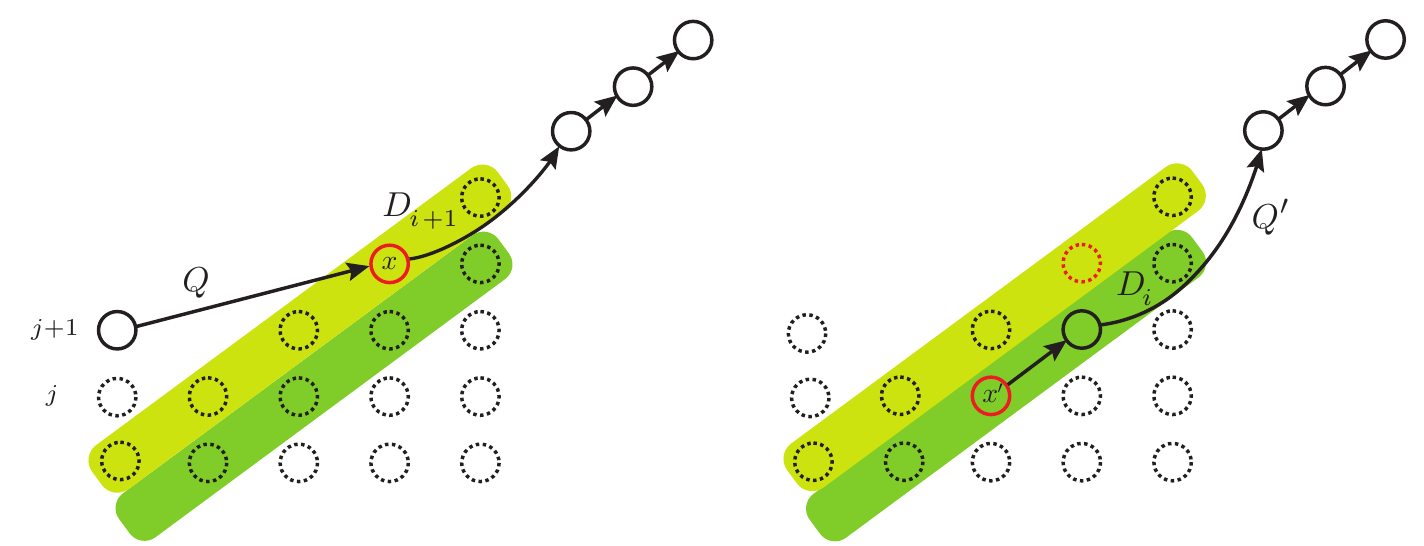}
\caption{A visualization of the proof idea of the central Claim in the proof of Lemma~\ref{lem:tensorlemma}.
The Claim asserts that if $D_{i+1}$ contains a vertex $x\in X$ with $\varphi(x)=j+1$, then
$D_i$ contains a vertex $x'\in X$ with $\varphi(x')\geq j$.
The upper diagonal $D_{i+1}$ is colored in light green, the lower diagonal $D_i$ is depicted in dark green.
We start by selecting a path $Q$ with $\text{lasthit}(Q) \in D_{i+1}$ and $\text{start}(Q)=j+1$.
This path is depicted on the left; the vertex $x=\text{lasthit}(Q)$ is highlighted in red.
We construct a path $Q'$ (shown on the right) such that $x'=\text{lasthit}(Q') \in D_{i}$ and $\text{start}(Q') = \text{start}(Q)-1$.
This path $Q'$ results from appending the end of path $Q$ to an appropriate subpath of the next lower diagonal $D_i$.
}
\label{fig:proof}
\end{figure*}

Now consider the following path $Q'$.
It begins with part of the diagonal $D_i$, namely $(v_{h+1-p},l(x)-p),\ldots,(v_h,l(x)-1)$,
and continues with the $k-p$ vertices from the part of $Q$ after $x$.
Note that by \eqref{eq:boundposition}
$$l(x)-p\geq \varphi(x)-1 \geq\max\{j,\text{start}(D_{i+1})-1\} \geq \max\{1,\text{start}(D_i)\},$$
so $Q'$ is well-defined.

The second part of $Q'$ does not contain any vertex from $X$ because $\text{lasthit}(Q)=x$.
Hence $x':= \text{lasthit}(Q')$ is in the diagonal part of $Q'$, i.e., in $D_i$.
By definition, $\varphi(x')\ge\text{start}(Q')=l(x)-p \ge j$.
\QED
\end{pf}

\section*{Conclusion}

We showed a simple $\frac{d}{2}$-approximation algorithm for (the deadline version of the discrete)
time-cost tradeoff problem, where $d$ is the depth. 
We used a reduction to the minimum-weight vertex cover problem in
$d$-partite hypergraphs and devised a deterministic algorithm that rounds a solution to the vertex cover LP. 
For this more general problem, 
it was known \cite{InapproxHypGraph} that no better approximation ratio is
possible, assuming the Unique Games Conjecture and $\textnormal{P}\not=\textnormal{NP}$. 
We proved that --- with the same assumptions --- no better approximation ratio than $\frac{d+2}{4}$ is possible for
time-cost tradeoff instances. 
Closing the gap between $\frac{d+2}{4}$ and $\frac{d}{2}$ remains an open problem.

\section*{Acknowledgements}
The authors thank Nikhil Bansal for fruitful discussions at an early stage of this project.

\end{document}